\documentclass[useAMS,usegraphicx]{mn2e}
\newcommand{\apj}{ApJ}
\newcommand{\apjl}{ApJL}
\newcommand{\mnras}{MNRAS}

\begin{document}
\title 
{On invisible plasma content in radio-loud AGNs:
The case of TeV blazar Markarian 421}

\author[]{M. Kino and F. Takahara\\
Department of Earth and Space Science,
Osaka University, Toyonaka 560-0043, Japan}

\author[M. Kino and F. Takahara]
{M. Kino$^{1,2}$ and
F. Takahara$^{3}$ \\
$^{1}$ ISAS/JAXA, 3-1-1 Yoshinodai, 229-8510 Sagamihara, Japan\\
$^{2}$ Department of Science $\&$ Engineering, Waseda University, 
169-8555 Tokyo, Japan\\
$^{3}$ Department of Earth and Space Science, Osaka University, 
560-0043 Toyonaka, Japan\\
}

\date{submitted to MNRAS ---- --- ---}

\maketitle

\begin{abstract}

Invisible plasma content in blazar jets
such as protons and/or thermal electron-positron ($e^{\pm}$) pairs 
is explored through combined arguments of 
dynamical and radiative processes. 
By comparing physical quantities required by the internal 
shock model with those obtained through the observed broadband 
spectra for Mrk 421, we obtain that
the ratio of the Lorentz factors of a pair of cold shells 
resides in about $2\sim 20$, which implies 
that the shocks are  at most mildly relativistic. 
Using the obtained Lorentz factors,
the total mass density $\rho$ 
in the shocked shells is investigated.
The upper limit of $\rho$ is obtained from the condition that 
thermal bremsstrahlung emission should not exceed  the observed 
$\gamma$-ray luminosity, 
whilst the lower limit is constrained from the 
condition that the energy density of 
non-thermal electrons is smaller than that of the total plasma.
Then we find $\rho$ is 
$10^2$-$10^3$ times heavier than that of non-thermal electrons
for pure $e^{\pm}$ pairs,
while  $10^2$-$10^6$ times heavier for pure electron-proton 
($e/p$) content, implying the existence of a large amount of
invisible plasma.
The origin of the continuous blazar sequence
is shortly discussed and we speculate that
the total mass density
and/or the blending ratio of $e^{\pm}$ pairs and $e/p$ plasma
could be new key quantities for the origin of the sequence.

\end{abstract}


\section{INTRODUCTION}

The discovery of strong inverse Compton components
in $X$ and $\gamma$-ray emission from jets in active galactic 
nuclei (hereafter AGN) for a wide range of spatial scales
(e.g., Collmar 2001 for review) enables us to probe 
quantitatively the energetics of relativistic jets. The 
kinetic power of non-thermal electrons has been estimated 
by various authors both for inner core jets (i.e., blazars)
(e.g., Kino, Takahara and Kusunose 2002, hereafter KTK;
Kusunose, Takahara and Kato 2003)
and large scale jets (e.g., 
Tavecchio et al. 2000; 
Leahy and Gizani 2001, 2002; 
Kataoka et al. 2003).
However, the material content of relativistic jets is not 
easily constrained by observations since the emission is 
dominated by that from non-thermal electrons and probably positrons
and it is difficult to directly constrain thermal matter 
content. 
Hence, the plasma composition in AGN jets, whether normal 
proton-electron ($e/p$) plasma 
or electron-positron pairs ($e^{\pm}$) is 
a dominant composition, is 
still a matter of open issue 
(e.g., Reynolds et al. 1996; 
Celotti, Kuncic, Rees and Wardle 1998; 
Wardle et al. 1999; Hirotani et al. 1999;
Sikora and Madejski 2000;
Ruszkowski and Begelman 2002;
Kino and Takahara 2004, hereafter KT04). 
This problem prevents us from estimating
the total mass and energy flux ejected from a
central engine.

To constrain invisible matter content such as thermal electron-positron
pairs and/or protons co-existing with non-thermal electrons, 
dynamical considerations are indispensable.
In KT04, we proposed a new procedure  
to constrain the invisible thermal plasma component
in classical FR II radio sources.
We used the fact that the mass and energy densities of 
the sum of thermal and non-thermal particles 
are larger than those of non-thermal electrons which 
are determined by observations.
Here we apply the same technique to the inner core jets of 
AGNs (i.e., blazars) 
based on the internal shock model.
The internal shock model is
believed to be most plausible 
to explain the production of high energy photons 
and time variabilities in blazars.
It has been widely applied also 
to the prompt emission of gamma-ray bursts (hereafter GRBs)
(e.g.,
Rees 1978; 
Rees and Meszaros 1994; 
Kobayashi, Piran and Sari 1997;,
Daigne and Mochkovitch 1998; 
Ghisellini 1999;
Spada, Ghisellini, Lazzati and Celotti 2001).
It is worth to note that
recently Ghisellini et al. (2005)
proposed a structured jet model cosisting of a fast spine 
surrounded by a slowly moving layer
for explaining VLBI scale radio blobs.
At the present, however, it is not evident where
is the acceleration site of electrons in the structured jet model. 
This is one of the prime issues which should be answered. 
Internal shocks are potentially the building blocks of 
the spine part of the structured jet. 
Whereas we recognize the importance of the
detailed structure of jets, 
as a first step 
we focus on the physical condition of the 
flow based on the simple internal shock model.

The methodology of constraining 
the invisible plasma content in the emission region is as follows.
As mentioned above,
a lower limit to the total mass density 
(sum of non-thermal electrons and invisible plasma)  
is restricted by the definition that the mass density
of total plasma should be smaller than that of 
the non-thermal electrons. 
The mass density of non-thermal electrons can be  
estimated by multi-frequency observations.
For this purpose, in \S \ref{sec:shock}
we review the shock dynamics of two colliding shells.     
Note that we do not use the simple
two point-mass approximation 
(e.g., Piran 1999; Lazzati et al. 1999;
Zhang and  M{\' e}sz{\' a}ros 2004 for review)
but employ the exact shock dynamics throughout this work.
This makes outcomes more accurate.
In \S \ref{sec:NT}
we briefly review the previous results 
on the amount of non-thermal electrons 
based on KTK.
In \S \ref{sec:invisible},
we constrain on the amount of total mass density.
As for the upper limit,
we use the constraint that bremsstrahlung emission from 
thermal electron (and positron) component should 
not exceed the observed $\gamma$-ray emission. 
We postulate synchrotron self-Compton (SSC) 
emission dominance in the $\gamma$-ray band which is
supported by the observed 
correlations between TeV$\gamma$-ray and X-ray 
in TeV blazars
(e.g., Takahashi et al. 1996, 2000; 
Catanese et al. 1997; Maraschi et al. 1999). 
We can thus bracket the amount of total mass density
in the emission region from below and above. 
In this way we apply this method to
 the archetypal TeV blazar Mrk 421.
In \S \ref{sec:dissipation},
we further estimate the shock dissipation rate of the 
colliding cold shells.
The dissipation rate
is a widely discussed quantity in literatures concerning gamma-ray bursts
(e.g., Lazzati, Ghisellini and Celotti 1999; Piran 1999).  
The shock dissipation is believed to be 
the ultimate source of heating and accelerating particles.
Summary and discussion are in 
\S \ref{sec:summary}.

\section{Key features of this work}

The key features of this work are
briefly summarised here in advance.
The existence of copious amount of invisible plasma 
is predicted from a qualitative consideration.

\subsection{Existence of invisible plasma content}

As mentioned in the introduction,
we constrain on the amount of invisible plasma content 
by introducing the dynamical considerations.
The point is that
we divide total mass and energy densities into 
two components, i. e., those of non-thermal electrons and those
of the other invisible components. 
The comparison of these obtained quantities enables
us to constrain on the amount of invisible plasma and
this is a new attempt compared with the previous works.

Bearing this in mind, 
next we show a quantitative consideration which derives
the existence of invisible plasma in colliding shells
of blazar jets.
Let us discuss a collision between 
a pair of equal mass-density shells for instance.
In the comoving frame of one shell, the particles of the other shell 
are coming in with a relative bulk Lorentz factor $\Gamma_{ij}$
(see in \S 3 for details) of a few at most (shown in Table 1). 
When only pair $e^{\pm}$ plasma are 
present and all of them are accelerated, then 
the average Lorentz factor of non-thermal electrons
$\langle\gamma_{e}\rangle$ is expected to be 
$\langle\gamma_{e}\rangle\approx\Gamma_{ij}$.
This is too small to account for observed blazar spectra 
 $\langle\gamma_{e}\rangle\approx 300$ (this is the case of Mrk 421) 
obtained by KTK. Therefore only a fraction of 
the pair $e^{\pm}$ should be accelerated, the ratio
of 
the rest mass density of 
non-thermal electrons 
to that of total plasma is about
$\Gamma_{ij}/\langle\gamma_{e}\rangle$.
Similarly we can discuss the case for shells with pure $e/p$ plasma makeup.  
If all of the dissipated energy goes into the electron acceleration, 
then we have $\langle\gamma_{e}\rangle \approx (m_{p}/m_{e})\Gamma_{ij}$. 
This is too large to account for the spectra
and it requires a limited fraction of electrons being accelerated.
Thus, invisible plasma is qualitatively expected when the 
internal shock is responsible for
the production of non-thermal electrons.
In this paper, we will quantitatively
explore the amount of invisible plasma in jets.

\subsection{Why we use shock dynamics?}

In the previous studies, 
the colliding shells have been approximately
modeled as the simple two-point-mass collision 
(e.g., Piran 1999; Lazzati et al. 1999;
Zhang and  M{\' e}sz{\' a}ros 2004).
The reason why we use the shock dynamics instead of the 
two-point-mass model is as follows.
When one try to derive the mass density from the mass,
one eventually needs to know lengths and velocities
and they can be consistently obtained by the shock model.
Hence the shock analysis is the best way
for investigating the invisible plasma content in jets.

\section{Shocks in colliding shells}\label{sec:shock}

Here we review the relativistic shock jump conditions.
We use one-dimensional shock dynamics 
of a pair of colliding shells to apply the 
standard internal shock model to blazars.  
Suppose the situation in which a rapid shell overtakes a 
previously ejected slow shell.
There are four characteristic regions designated by
(1) unshocked slow shell,
(2) shocked slow shell,
(3) shocked rapid shell, and
(4) unshocked rapid shell.
These regions are separated by 
the forward shock (FS),
the contact discontinuity (CD), and
the reverse shock (RS).
In this paper, we use the terminology 
of  {\it regions} $i$ ($i$=1, 2, 3, and 4) and
{\it position of discontinuity}
$i$ ($i$=FS, CD, and RS) where 
FS, CD, and RS stand for the forward shock front, 
contact discontinuity,
and reverse shock front, respectively.
The fluid velocity and Lorentz factor in the region $i$ 
measured in the interstellar medium (hereafter ISM) frame  
are expressed as
$v_{i}(=\beta_{i}c)$ and $\Gamma_{i}$, respectively.
The relative velocity and Lorentz factor of the fluid $i$ 
measured in the frame $j$ are denoted by 
$v_{ij}(=-v_{ji}=\beta_{ij}c=-\beta_{ji}c)$ 
and $\Gamma_{ij}(=\Gamma_{ji})$, respectively. 
Rest mass density, pressure, and internal energy density are 
expressed as
$\rho_{i}$,
$P_{i}$, and 
$e_{i}$,
respectively. As for the equation of state (EOS), we take
$P_{i}=(\hat{\gamma}_{i}-1)(e_{i}-\rho_{i}c^2)$,
where $\hat{\gamma}_{i}$ is the adiabatic index.
We sometimes use the subscripts s and r 
instead of $1$ and $4$, such as 
$\Gamma_{1}=\Gamma_{\rm s}$ and 
$\Gamma_{4}=\Gamma_{\rm r}$.

In the limit of strong shock, 
with the assumption of cold upstream ($P_1=0$),
the jump conditions for the forward shock 
are written as follows
(Blandford \& McKee 1976):
\begin{eqnarray}\label{eq:FS}
\Gamma_{\rm FS1}^{2}
=\frac{(\Gamma_{\rm 12}+1)[\hat{\gamma}_{2}(\Gamma_{\rm 12}-1)+1]^{2}}
{\hat{\gamma}_{2}(2-\hat{\gamma}_{2})(\Gamma_{\rm 12}-1)+2},\nonumber \\
e_{2}=\Gamma_{\rm 12}\rho_{2} \, ,
\qquad
\frac{\rho_{2}}{\rho_{1}}=
\frac{\hat{\gamma}_{2}\Gamma_{12}+1}{\hat{\gamma}_{2}-1} \, ,
\end{eqnarray}
where $\Gamma_{12}=\Gamma_{1}\Gamma_{2}(1-\beta_{1}\beta_{2})$, and
$\Gamma_{\rm FS1}$
is the Lorentz factor of forward shock measured 
in the rest frame of the unshocked slow shell. 
In the relativistic limit, the adiabatic index is 
${\hat \gamma}_{2}=4/3$.  
Using the same assumptions as in the forward shock,
the jump conditions for the reverse shock are given by:
\begin{eqnarray}\label{eq:RS}
\Gamma_{\rm RS4}^{2}
=\frac{(\Gamma_{\rm 34}+1)[\hat{\gamma}_{3}(\Gamma_{\rm 34}-1)+1]^{2}}
      {\hat{\gamma}_{3}(2-\hat{\gamma}_{3})(\Gamma_{\rm 34}-1)+2},\nonumber \\
e_{3}=\Gamma_{\rm 34}\rho_{3} \, ,
\qquad
\frac{\rho_{3}}{\rho_{4}}=
\frac{\hat{\gamma}_{3}\Gamma_{34}+1}{\hat{\gamma}_{3}-1} \, ,
\end{eqnarray}
where $\Gamma_{34}=\Gamma_{3}\Gamma_{4}(1-\beta_{3}\beta_{4})$,
and $\Gamma_{\rm RS4}$
is the Lorentz factor of the reverse shock measured in 
the rest frame of the unshocked rapid shell. 
The equality of pressure and velocity across the contact discontinuity 
gives
\begin{eqnarray}
P_{2}=P_{3} ,
\qquad
\Gamma_{2}=\Gamma_{3} \, .
\end{eqnarray}

After the shocks break out the shells,
$\Gamma_{2}=\Gamma_{3}$ is not satisfied
because a rarefaction wave changes the 
density and velocity profiles 
(e.g., Kino, Mizuta and Yamada 2004, hereafter KMY).
We do not treat the rarefaction waves for simplicity, 
concentrating on the major duration before shock 
breakout.
It may be useful 
to rewrite the pressure balance along the CD as
\begin{eqnarray}\label{eq:ratiorho}
\frac{\rho_{4}}{\rho_{1}}
=\frac{(\hat{\gamma}_{2}\Gamma_{12}+1)(\Gamma_{12}-1)}
{(\hat{\gamma}_{3}\Gamma_{34}+1)(\Gamma_{34}-1)}  \ .
\end{eqnarray}
%

In general, 
the number of physical quantities
in each region is $3$, 
$\rho_{i}$, $P_{i}$ (or $e_{i}$), and $v_{i}$.
Forward and reverse shock speeds 
(i.e., $v_{\rm FS}$ and $v_{\rm FS}$) 
are two other quantities.
In all, there are $3\times 4+2=14$ physical quantities. 
Note that $P_i$ and $e_i$ are connected with EOS. 
The total number of the jump conditions 
is $3+3+2=8$.
Hence, given $3+3=6$ upstream quantities 
for each shock, 
we can obtain the remaining $8$ downstream 
quantities by using $8$
jump conditions. It is to be noted that the absolute value of 
the rest mass density is irrelevant to the shock dynamics 
since the shock dynamics is linear with respect to the mass desnity. 
Then, actually we need to specify 5 quantities if we 
give the density ratio $\rho_4/\rho_1$.


For a specific case for TeV blazars,
we here impose the following two conditions;
(i) the unshocked shells are cold, 
i.e., $P_{1}=P_{4}=0$, 
(ii) the Lorentz factor of the 
shocked regions $\Gamma_{3}(=\Gamma_{2})$ 
is identified as that of the 
emission region obtained by the  
observed broadband spectra. 
Further, we examine the following 
three cases for the ratio $\rho_{\rm r}/\rho_{\rm s}$; 
(a) the energy of bulk motion of the rapid shell 
($E=\Gamma mc^{2}$) 
equals to that of the slow one in the ISM frame
(we refer to it as ``equal energy (or $E$) case''),
(b) the mass of the rapid shell ($m=\rho \Gamma\Delta$)
equals to that of the slow one 
(hereafter we call it ``equal mass (or $m$) case''), and 
(c) the rest mass density of the rapid shell equals to that of the slow one 
(hereafter we call it ``equal rest mass density (or $\rho$) case'').
Here, $\Delta$ denotes the thickness of the shell 
measured in ISM frame. 
These choices are based on the conjecture that 
the ejecta from the ``central engine'' is likely 
to have a correlation with each other (e.g., NP02; KMY).
Hereafter, we assume that the widths of two shells are the 
same in the ISM frame, that is $\Delta_{\rm r}/\Delta_{\rm s}=1$ 
(e.g., NP02, Spada et al. 2001).
Note that in the case of 
$\Delta_{\rm r}=\Delta_{\rm s}$
and $\Gamma_{\rm r} >\Gamma_{\rm s}$,
$\rho_{\rm s}$ is always
larger than  $\rho_{\rm r}$ for equal $E$ and equal $m$ cases.

Thus, we give $4$ quantities, $P_1$, $P_4$, $\Gamma_2=\Gamma_3$
and one relation between the rapid and slow shells, 
$\rho_4/\rho_1$ depending on cases (a) 
through (c) described above.
As a remaining quantity, the Lorentz factor of the 
rapid shell
$\Gamma_{4}$  is treated as a free parameter.
Although we do not specify the absolute value of $\rho$, 
we treat the abosolute value in actual applications. 
It is compared with 
that of non-thermal electrons in the shocked regions as described 
in \S \ref{sec:invisible}. 
The absolute value of the rest mass density comes into play 
when two-body processes 
such as bremsstrahlung emission is used to obtain the upper limit 
of $\rho$.
We will properly discuss these points.

In the following sections,
we focus on the values of 
(i)  the value of $\Gamma_{1}$ and $\Gamma_{4}$,
(ii) $e_{3}$ and/or $\rho_{3}$,
as a tool to examine the physical quantities of 
invisible matter content.

\section{Amount of non-thermal electrons}\label{sec:NT}

\subsection{Number and energy densities}

Based on the detection of inverse Compton emission in 
$\gamma$-ray band,
the number and energy densities of 
the non-thermal (hereafter ``NT'')
electrons $n_{e}^{\rm NT}$ and $e_{e}^{\rm NT}$  
in shocked regions can be determined
by the comparison of 
the observed broadband spectrum and 
the theoretical one. Although the minimum Lorentz factor 
of relativistic electrons 
is not definitely determined and affects mainly 
the number density $n_{e}^{\rm NT}$, 
we regard that low energy electrons below $\gamma_{e,\rm min}$ 
constitute thermal electrons. 
Considering the observed flat number spectrum of electrons,  
fixing $\gamma_{e,\rm min}=10$ does not cause any major problem with 
$n_{e}^{\rm NT}$.  

Here, we briefly quote the resultant 
$n_{e}^{\rm NT}$ and $e_{e}^{\rm NT}$
obtained in KTK.
Hereafter, 
we omit the subscript expressing the 
regions $i(=2,3)$ for simplicity.
For clearness of the following argument,
we define that $n_{e}^{\rm NT}$ and $e_{e}^{\rm NT}$
also include NT positrons when they exist. 
The quantity $n_{e}^{\rm NT}$  is written as
$n_{e}^{\rm NT}\equiv
\int^{\infty}_{\gamma_{e,\rm min}}
n_{e}(\gamma_{e})d\gamma_{e}$,
while $e_{e}^{\rm NT}$ is given by
$e_{e}^{\rm NT}=
\langle\gamma_{e}\rangle
n_{e}^{\rm NT}m_{e}c^{2}$, 
where 
$n_{e}(\gamma_{e})$ and 
$\langle\gamma_{e}\rangle$ are the energy spectrum and  
the average Lorentz factor of NT electrons, respectively.
By a detailed comparison of the SSC model with
observed broadband spectrum of Mrk 421,
we obtained $n_{e}^{\rm NT}$ as
\begin{eqnarray} \label{eq:??}
n_{e}^{\rm NT}
\simeq
11\times
\left(
\frac{\gamma_{e,\rm min}}{10}
\right)^{-0.6} \ \rm cm^{-3}.
\end{eqnarray}
Here, we adopt the index of injected electrons for Mrk 421 
as $s=1.6$ 
(e.g., Mastichiadis \& Kirk 1997; Kirk \& Duffy 1999)
and the case of $\gamma_{e,\rm min}=10$ was examined in KTK.
The best choice of the size of the emission region is 
$R=2.8 \times 10^{16}{\rm cm}$ with an order of magnitude 
uncertainty. Thus, the corresponding uncertainty of $n_{e}^{\rm NT}$ 
amounts to two orders of magnitude; for smaller $R$, larger 
$n_{e}^{\rm NT}$ should be adopted. 
But, as far as the the shock dynamics is concerned, only 
the density ratio plays a role, therefore we adopt the above value 
as the canonical one.  

As for the average energy of NT electrons, we obtained  
\begin{eqnarray} \label{eq:uacc}
e_{e}^{\rm NT}/n_{e}^{\rm NT}
=\langle\gamma_{e}\rangle m_{e}c^{2}
\simeq
3.1\times 10^{2} m_{e}c^{2}.
\end{eqnarray}
Since for $s=1.6$,  
electrons near the cooling break energy $\sim \gamma_{e,\rm br}$
carry most part of the kinetic energy
and $e_{e}^{\rm NT}$ has a weak dependence on $\gamma_{e,\rm min}$
provided that $\gamma_{e,\rm min}$ is smaller than
$\gamma_{e,\rm br}\sim 10^{4}$.
Note that the case of 
$\gamma_{e,\rm min}\sim 10^{4}$ is ruled out for Mrk 421
since the case does not fit the EGRET data (KTK).
Therefore, Eq. (\ref{eq:uacc}) is justified in any case 
for Mrk 421.


\subsection{Forward and reverse shocks}

To clarify whether the observed non-thermal 
emission comes mainly from FS or from RS region, 
the typical frequency of non-thermal synchrotron radiation 
and internal energy density in each region are examined here.

According to the standard diffusive shock acceleration, 
the acceleration time scale is estimated as (e.g., Drury 1983) 
$t_{\rm acc}=(2\pi\gamma_{e} m_{e}c\xi)/(eB)$ 
where 
$\xi=\lambda/r_{g}$
is a parameter related to the amplitude of 
magnetic fluctuations,  $\lambda$ and $r_{g}$ are
the mean free path for 
the scattering of electrons and  Larmor radius, respectively.
Here the shock speed is taken to be $c$. 
On the other hand,
the synchrotron cooling time is given by
$t_{\rm syn}=
(6\pi m_{e}c^{2})/(\sigma_{T}\gamma_{e}cB^{2})$.
The maximum Lorentz factor of 
the non-thermal electrons is evaluated as
$\gamma_{\rm max}\propto B^{-1/2}$
by using the condition of $t_{\rm acc}=t_{\rm syn}$ 
at $\gamma_{\rm max}$
with the assumption 
that $\xi$  in FS and RS regions takes the same value. 
Hence,
the characteristic synchrotron photon energy is given by
$h\nu_{\rm syn,o,max}
\propto \Gamma_{i}B\gamma_{\rm max}^{2}={\rm const.}$,
Hence, the value $\nu_{\rm syn,o,max}$ 
in FS region and RS regions is the same. 


The total internal energy of NT electrons
in FS and RS regions may be discussed as follows.
If $\Gamma_{21}\gg1$ and $\Gamma_{43}\gg1$
are satisfied, we have
$e_{2}+P_{2}\simeq e_{3}+P_{3}$. In the actual case of blazars,
$\Gamma_{ij}$ is close to order of unity and we have  
$P_{2}=P_{3}$.
Thus, 
the energy densities of regions 2 and 3 are similar and 
the internal energy 
is controlled by the comoving shell widths.
Since $\Gamma_{43}\ge\Gamma_{21}$ is always 
satisfied, co-moving length of RS region is 
larger than that of FS region  
in the case of $\Delta_{\rm r}=\Delta_{\rm s}$ 
(e.g., Kobayashi and Sari 2001; NP02; KMY).  
Thus, the radiation from RS dominates over that from FS region. 
Based on this consideration, 
we focus on RS dominated case in this paper.
Hereafter, we omit the subscript $3$ for simplicity.

\section{Constraints on the amount of invisible plasma}\label{sec:invisible}


\subsection{Lorentz factors of cold shells}\label{subsec:coldshell}

It is hard to estimate the 
bulk Lorentz factors of cold shells
simply because they are invisible. 
However, by using the value of Lorentz factor of shocked shell  which
corresponds to the beaming factor of the emission region,  
we can constrain on the 
Lorentz factors of the cold shells.
Here, we consider the range from 3 to 100 for $\Gamma_{\rm r}$
and $\Gamma_{\rm s}$. 
Following   
Begelman, Rees \& Sikora (1994),
we consider the upper limit of the 
Lorentz factors of the emission region as 
$\Gamma_{\rm r,max}= 100$,
while
as for the lower limit
we employ $\Gamma_{\rm s,min}=3$ 
based on Wardle \& Aaron (1997). 
Here, we exclude cases of very
weak collisions with 
$\Gamma_{\rm r}/\Gamma_{\rm s}<2$
as in NP02. 
As for the adiabatic index in Eq. (\ref{eq:ratiorho}),
we approximate $\hat{\gamma}_{i}=4/3$ for $\Gamma_{ij}>2$, otherwise
$\hat{\gamma}_{i}=5/3$ for simplicity (e.g., Kirk and Duffy 1999).

For the TeV blazar Mrk 421, we have already 
obtained $\Gamma_2=\Gamma_{3}=12$ by the observed
multi-frequency spectrum (KTK). 
Hence, 
Eq. (\ref{eq:ratiorho}) is solvable for
$\Gamma_{\rm s}$ given $\rho_{\rm r}/\rho_{\rm s}$ 
and $\Gamma_{\rm r}$. 
%
%
Qualitatively, a faster $\Gamma_{\rm r}$ requires a slower 
$\Gamma_{\rm s}$ to attain the
same value of $\Gamma_{3}$.
Thus, minimun value of $\Gamma_{\rm r}$ corresponds to 
$\Gamma_{\rm r}/\Gamma_{\rm s}=2$, while the maximum value of
$\Gamma_{\rm r}$ corresponds to $\Gamma_{\rm r}=100$ 
or $\Gamma_{\rm s}=3$.

In  Table \ref{table:1},
we show 
the minimum and maximum values of  
$\Gamma_{\rm r}$ and  $\Gamma_{\rm s}$ and 
the corresponding relative Lorentz factors 
$\Gamma_{12}$ and
$\Gamma_{34}$  which control the shock heating of the downstreams 
(see Eqs. (\ref{eq:FS}) and (\ref{eq:RS})).
From this, we see that
the range of $\Gamma_{ij}$ lies between 1.03 and 4.2.  
In other words, a mildly relativistic shock is realized
in the case of Mrk 421.
We also note that our adopted value of $\gamma_{e,\rm min}=10$ 
is a reasonable choice with the assumption that 
$\gamma_{e,\rm min}$ should be a few times larger than $\Gamma_{ij}$.
The corresponding value of  
$\Gamma_{\rm r}/\Gamma_{\rm s}$ is found as
\begin{eqnarray} 
2<\Gamma_{\rm r}/\Gamma_{\rm s}\le 16.0  &&   
({\rm equal} \ \rho),  \nonumber \\
2<\Gamma_{\rm r}/\Gamma_{\rm s}\le 19.5 &&   
({\rm equal} \ m), ~{\rm and} \nonumber \\
2<\Gamma_{\rm r}/\Gamma_{\rm s}\le 11.7 &&   
({\rm equal} \ E)  ,
\end{eqnarray} 
respectively.

\begin{table}
\caption{Lorentz factors obtained
by internal shock analysis for Mrk 421}
\label{table:1}
\begin{tabular}{l| c r r r }
\hline\hline
case 
& $\Gamma_{\rm s}$  
& $\Gamma_{\rm r}$ 
& $\Gamma_{12}$
& $\Gamma_{43}$    \\
\hline
equal $\rho$ (largest $\Gamma_{\rm r}/\Gamma_{\rm s}$)
&  3
&  48.0
& 2.125
& 2.125
\\
equal $\rho$ (smallest $\Gamma_{\rm r}/\Gamma_{\rm s}$)
&  8.485
&  16.97
& 1.060
& 1.060
\\
equal $m$ (largest $\Gamma_{\rm r}/\Gamma_{\rm s}$)
& 5.12
& 100
& 1.35
& 4.22
\\
equal $m$ (smallest $\Gamma_{\rm r}/\Gamma_{\rm s}$)
& 8.983
& 17.959
& 1.042
& 1.082
\\
equal $E$ (largest $\Gamma_{\rm r}/\Gamma_{\rm s}$)
&  8.57
&  100
& 1.057
& 4.226
\\
equal $E$ (smallest $\Gamma_{\rm r}/\Gamma_{\rm s}$)
&  9.48
&  18.96
&  1.027
&  1.106
\\
\hline
\end{tabular}\\
Notes: 
$\Gamma_{\rm max}=100$, 
$\Gamma_{\rm min}=3$,
 $\Gamma_{\rm r}/\Gamma_{\rm s}>2$, and  
$\Gamma_{2}=\Gamma_{3}=12$, are employed
in this analysis.
\end{table}

\subsection{Total mass density}\label{subsec:massdensity}

\subsubsection{Lower limit of $\rho$}\label{sec:lower-rho}

In \ref{subsec:coldshell}, we show that
shock is at most mildly relativistic
though each shell moves at a relativistic speed.
As a consequence, dissipation efficiency 
is relatively small and
$\langle\gamma_{e}\rangle \gg \Gamma_{34}$ is satisfied.
Therefore $e_{e}^{\rm NT}/e
=\langle\gamma_{e}\rangle\rho_{e}^{\rm NT}/
\Gamma_{34}\rho<1$ gives a tighter constraint than 
$\rho_{e}^{\rm NT}/\rho<1$.
By rewriting the condition of 
$e_{e}^{\rm NT}/e<1$, the lower limit of 
the total mass density is given by 
\begin{eqnarray}\label{eq:lower}
\frac
{\rho} 
{\rho_{e}^{\rm NT}}
&>& \frac
{\langle\gamma_{e}\rangle}
{\Gamma_{34}}
\simeq \frac{3.1\times 10^{2}}{\Gamma_{34}} .
\end{eqnarray}
Here we omit the subscript of region number $i=3$
for the various densities for thumbnail writing.
From this 
we directly see that
in order to accelerate electrons up to 
$\langle\gamma_{e}\rangle\sim 3.1\times10^{2}$ in the framework of standard
internal shock model, where only 
a small available shock dissipation energy   
$\Gamma_{34}\sim a \ few$ is realized,
the invisible mass density at least about 100 times the rest mass density 
of NT electrons is definitely required. 
In other words,
we need a loading of baryons and/or a thermal pair plasma. 
It is worth to note the effects of an uncertainty with
$\langle\gamma_{e}\rangle$. 
We estimated the uncertainty range as 
$ 2.3\times 10^{2}
< \langle\gamma_{e}\rangle <
4.3\times 10^{2}$ (KTK).
The uncertainty simply leads to a shift of 
lower limit curve by the same factor.
Since it causes only a small change on the resultant value,
we focus on the best-fit case in this work  
for simplicity.

\subsubsection{Upper limit of $n_{e}^{\rm T}$}\label{sec:upper-n}

Here, we constrain the upper limit of the number density of 
thermal electrons $n_{e}^{\rm T}$.
As mentioned in the Introduction,
it is widely accepted that
observed GeV and TeV $\gamma$-rays are  
SSC dominated.
In the MeV range, bremsstrahlung radiation by the
thermal electrons with temperature 
$\Theta_{e}\equiv kT_{e}/m_{e}c^{2}
\sim \Gamma_{34} \sim $ a few MeV  
is expected 
if adequate amount of thermal electrons exist in the
emission region.
At the moment,
we do not have any observational evidence for 
the bremsstrahlung in MeV band.
At the same time, it is fair to note that
observation in MeV  range itself is a challenging area
(e.g., Takahashi et al. 2003).
Here we estimate the upper limit of the number density of
thermal electrons
by assuming the observed bolometric luminosity of
bremsstrahlung $L_{{\rm brem,o}}$ should be lower 
than that of SSC $L_{\rm ssc,o}$ which is estimated as
$ L_{{\rm ssc,o}}=
7 \times 10^{44}$ erg s$^{-1}$ (KTK).

For $e^{+}$$e^{-}$ plasma content,
we employ 
Eqs. (21) and (22)
of Svensson (1982) which express
the emissivity of relativistic $e^{+}$$e^{-}$ 
bremsstrahlung $\epsilon_{\rm brem}(\Theta_{e},n_{e}^{\rm T})$
where $n_{e}^{\rm T}$ is
the number density of thermal electrons.
Note that these expressions do not include the bremsstrahlung
between electron-electron and positron-positron
and the limit will be severer by a factor of $\sim 2$
if we include them.
Then, the condition of 
$ L_{\rm ssc,o}>L_{\rm brem,o}$
is rewritten as
\begin{eqnarray}\label{eq:upper1}
n_{e}^{\rm T}
&<&  9.7 \times 10^{2} 
\left[\Theta_{e}^{1/2}(1+1.7\Theta_{e}^{1.5})
\right]^{-1/2} 
\rm \ cm^{-3} \quad (\Theta_{e}<1) \nonumber \\
&<&  5.7 \times 10^{2} 
\left[
\Theta_{e}(\ln(1.1\Theta_{e})+5/4
\right]^{-1/2} 
\rm \ cm^{-3}\quad (\Theta_{e}\ge1) .  \nonumber \\
\end{eqnarray}
The 
bolometric luminosity
of the optically-thin bremsstrahlung is estimated by
$L_{{\rm brem,o}}
=(4\pi R^{3}/3) \Gamma_{3}^{4} 
\epsilon_{\rm brem}$ 
with the emission size 
$R=2.8\times 10^{16}\rm cm$ 
and the Lorentz factor
$\Gamma_{3}=12$ as obtained by the broadband spectral fitting
of Mrk 421 (KTK).
The electron temperature is evaluated by  
$(\hat{\gamma}_3-1)\Theta_{e}=\Gamma_{34}-1$.
The upper limit turns out to be about a thousand times 
larger than the mass density of non-thermal electrons. 
It is consistent with and relatively close to
the required lower limit of the
mass density by Eq. (\ref{eq:lower}).
This upper limit depends on the adopted value of $R$, 
and it is proportional to $R^{-3/2}$. 
Considering that $n_e^{\rm NT}$ is roughly proportional 
to $R^{-2}$, the ratio of this upper limit to $n_e^{\rm NT}$
only has a weak dependence on $R$.

Similarly, in the case of electron-proton 
(hereafter $e/p$) plasma content,
we can rewrite the condition of 
$ L_{\rm ssc,o}>L_{\rm brem,o}$
as
\begin{eqnarray}\label{eq:upper2}
n_{e}^{\rm T}
&<&  9.5 \times 10^{2} 
\left[\Theta_{e}^{1/2}(1+1.78\Theta_{e}^{1.34})
\right]^{-1/2} \rm \ cm^{-3} 
   (\Theta_{e}<1) \nonumber \\
&<&  9.6 \times 10^{2} 
\left[
\Theta_{e}(\ln(1.1\Theta_{e}+0.42)+3/2
\right]^{-1/2}    
\rm \ cm^{-3}  \nonumber  \\ 
&& (\Theta_{e}\ge1) 
\end{eqnarray}
with  Eqs. (17) and (18)
of Svensson (1982).
Note that electron-electron 
bremsstrahlung is not considered in these equations.
It is clear that the 
upper limit of $\rho$ in this case is 
$m_{p}/m_{e}$ times larger than $n_{e}^{\rm T}m_e$.

Lastly, let us check 
the timescale of  $e^{\pm}$ pair 
annihilation $t_{\rm ann}$.
It is evaluated as $t_{\rm ann}
\simeq \Theta_{e}^{2}/(n_{e}\sigma_{T}c)
\simeq 6 \times 10^{10}\Theta_{e}^{2}
(n_{e} /10^{3} \rm cm^{-3})^{-1}$ sec.
Hence we see that
the annihilation time scale is 
much longer than the dynamical time scale 
$t_{\rm dyn}\equiv \sqrt{3}R/c
\approx 2\times 10^{6}(R/10^{16}~{\rm cm})~{\rm sec}$.
Therefore  $e^{\pm}$ pair 
annihilation is not effective in this situation.

\subsubsection{Allowed range of $\rho$}\label{sec:allowed}

\begin{figure} 
\includegraphics[width=8cm]{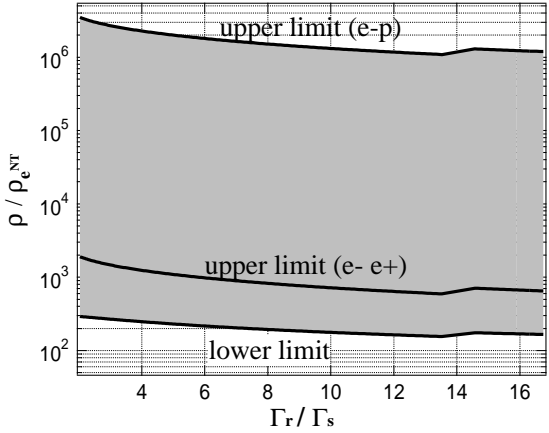}
\caption
{The allowed region of  
the amount of mass density of total plasma
normalized by that of non-thermal electrons
$\rho/\rho_{e}^{\rm NT}$. The gray region is allowed for 
pure pair plasma 
for ``equal $\rho$'' case. The line denoted by $e/p$ 
shows the upper limit for $e/p$ plasma.
Horizontal axis shows the ratio of the Lorentz factor
of a rapid shell to a slow one which lies in 
$2<\Gamma_{\rm r}/\Gamma_{\rm s}<16.0$.}
\label{fig:eqrho}
\end{figure}

\begin{figure} 
\includegraphics[width=8cm]{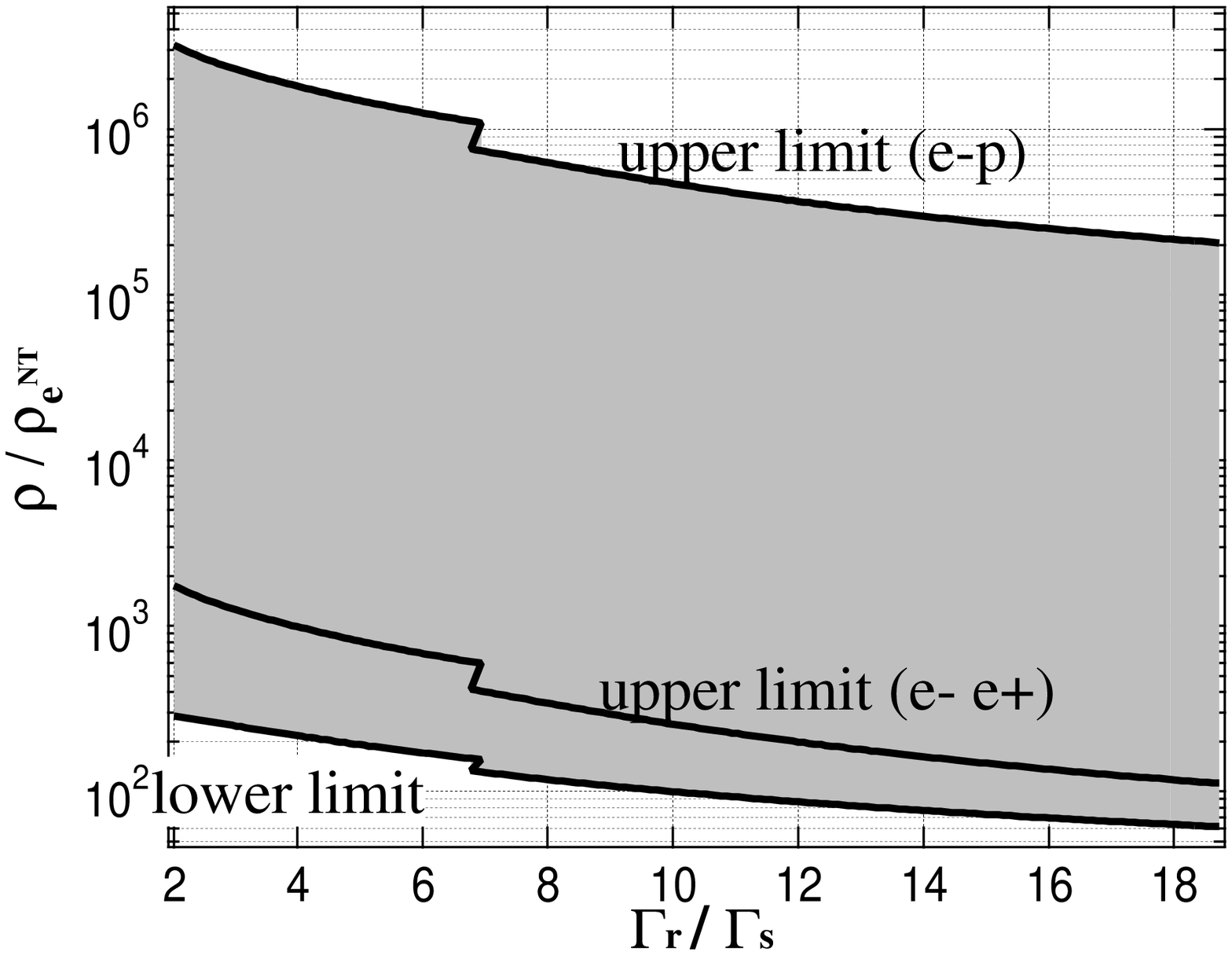}
\caption
{The allowed region of 
the amount of mass density of total plasma
normalized by that of non-thermal electrons
$\rho/\rho_{e}^{\rm NT}$. The  gray region is allowed 
for pur pair plasma 
for ``equal $m$'' case. The line denoted by $e/p$ 
shows the upper limit for $e/p$ plasma.
Horizontal axis shows the ratio of the Lorentz factor
of a rapid shell to a slow one which lies in 
$2<\Gamma_{\rm r}/\Gamma_{\rm s}<19.5$. }
\label{fig:eqmass}
\end{figure}

\begin{figure} 
\includegraphics[width=8cm]{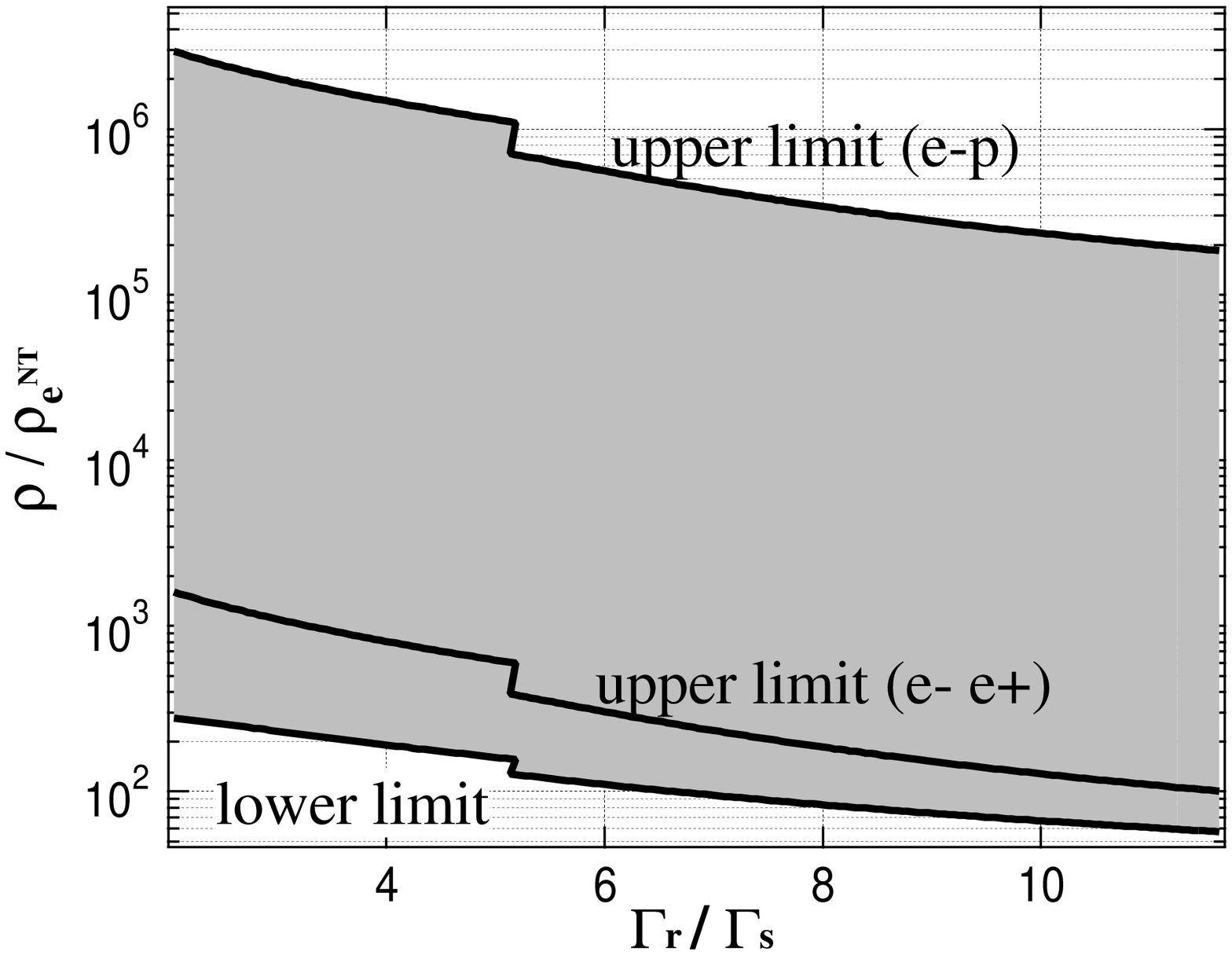}
\caption
{The allowed region of 
the amount of mass density of total plasma
normalized by that of non-thermal electrons
$\rho/\rho_{e}^{\rm NT}$. the gray region is allowed 
for pure pair plasma  
for ``equal $E$'' case. The line denoted by $e/p$ 
shows the upper limit for $e/p$ plasma.
Horizontal axis shows the ratio of the Lorentz factor
of a rapid shell to a slow one which lies in 
$2<\Gamma_{\rm r}/\Gamma_{\rm s}<11.7$. }
\label{fig:eqE}
\end{figure}

We thus obtained the upper and lower limits on 
$\rho/\rho_e^{\rm NT}$ and the results are shown 
in the plane of mass density of invisible plasma and 
$\Gamma_{\rm r}/\Gamma_{\rm s}$
in the cases of 
``equal $\rho$'', 
``equal $m$'', and
``equal  $E$''
in Figs. 
\ref{fig:eqrho},
\ref{fig:eqmass}, and
\ref{fig:eqE}, respectively.
They are obtained by solving Eq. (\ref{eq:ratiorho})
and inserting $\Gamma_{34}$
into Eqs. (\ref{eq:lower}), (\ref{eq:upper1}), and (\ref{eq:upper2}).
The qualitative features are the same for these three cases, 
although different in quantitative detail.
Summing up in advance, the most important result is that a
large amount of mass density of invisible plasma 
is required in the emission region.
As the value of $\Gamma_{r}/\Gamma_{s}$ increases, 
the value of $\Gamma_{34}$ becomes larger
and the lower limit on the invisible mass density 
($\rho/\rho^{\rm NT}_{e}$) 
reduces. 
Below we discuss two extreme cases of different plasma content.
One is the case of the jet 
with pure $e^{\pm}$ pair plasma content,
whilst
the other is the jet made of pure $e/p$ plasma.

For pure $e^{\pm}$ pair jet,
the resultant total mass density 
normalized by $\rho_{e}^{\rm NT}$ is 
\begin{eqnarray}\label{eq:e-rho}
2\times 10^{2}<\rho
/\rho^{\rm NT}_{e}<2\times 10^{3} && ({\rm equal} ~ \rho) \nonumber \\ 
7\times 10^{1}<\rho
/\rho^{\rm NT}_{e}<2\times 10^{3} && ({\rm equal} ~ m)\nonumber \\ 
6\times 10^{1}<\rho
/\rho^{\rm NT}_{e}<2\times 10^{3} && ({\rm equal} ~ E)   .
\end{eqnarray}
For the jets consisting of pure $e^{\pm}$ plasma, 
the predicted $\rho/\rho^{\rm NT}_{e}$ is 
constrained in a narrow range around 100-1000
as shown in Figs. \ref{fig:eqrho},
\ref{fig:eqmass}, and \ref{fig:eqE}. 
The number density fractions of the shock accelerated 
$e^{\pm}$ pairs are directly obtained as 
$
\rho^{\rm NT}_{e}/\rho=n^{\rm NT}_{e}/(n^{\rm T}_{e}+n^{\rm NT}_{e})
\sim
10^{-3}-10^{-2}$.
This seems a reasonable result since the number of accelerated 
particles is expected to be a small fraction of the thermal pool.

In the case of pure $e/p$ content, the allowed range
of $\rho/\rho^{\rm NT}_{e}$
are found to be
\begin{eqnarray}\label{eq:p-rho}
2\times 10^{2}<\rho
/\rho^{\rm NT}_{e}<3\times 10^{6} && ({\rm equal} ~ \rho), \nonumber \\ 
7\times 10^{1}<\rho
/\rho^{\rm NT}_{e}<3\times 10^{6} && ({\rm equal} ~ m), ~{\rm and} \nonumber \\ 
6\times 10^{1}<\rho
/\rho^{\rm NT}_{e}<3\times 10^{6} && ({\rm equal} ~ E),
\end{eqnarray}
respectively.
The maximum values
of $\rho/\rho^{\rm NT}_{e}$ are about $m_{p}/m_{e}$
times larger than those 
in the case of pure $e^{\pm}$ pair content.

%
%

\subsection{Allowed range of $e/e_{e}^{\rm NT}$}

As shown above, the lower and upper limit of 
$\rho/\rho^{\rm NT}_{e}$ have been obtained in \ref{sec:lower-rho}
and \ref{sec:upper-n}, respectively.
By using the obtained $\rho/\rho^{\rm NT}_{e}$ 
shown in \ref{sec:allowed}, 
we can estimate 
$e/e^{\rm NT}_{e}=\Gamma_{34}\rho/
\langle\gamma_{e}\rangle\rho^{\rm NT}_{e}$.
For the case of pure $e^{\pm}$ content,
since the allowed range of $\rho/\rho^{\rm NT}_{e}$ 
is narrow, the corresponding
$e/e^{\rm NT}_{e}$ is also well
constrained as
\begin{eqnarray}
1<e
/e^{\rm NT}_{e}< 7 && ({\rm equal} ~ \rho), \nonumber \\ 
1<e
/e^{\rm NT}_{e}< 7 && ({\rm equal} ~ m), ~{\rm and } \nonumber \\ 
1<e
/e^{\rm NT}_{e}< 7 && ({\rm equal} ~ E),
\end{eqnarray}
where we employ Eq. (\ref{eq:e-rho}) and Table 1.
Thus we find that
$e/e^{\rm NT}_{e}\le
2\times 10^{3}\times 1.1/310 \approx 7$.
In other words, for $e^{\pm}$ pair content,
the total kinetic power of  the shocked (emission) 
region is less than 
$L_{\rm kin}\approx 7L_{\rm kin,e}^{\rm NT}$
where $L_{\rm kin,e}^{\rm NT}$ is the kinetic power of 
NT electrons estimated as
$L_{\rm kin,e}^{\rm NT}=
4\times 10^{44}
$erg s$^{-1}$ (KTK). 
In the case of $e/e^{\rm NT}_{e}\approx 1$, 
the non-linear dynamical structure of the shock 
(e.g., Drury and Voelk 1981; 
Berezhko and Ellison 1999)
is required for analysing the phenomena
at the vicinity of  the shock front.
Note that the case discussed here 
is consistent with our choice of $\Gamma_{r}/\Gamma_{s}>2$.

On the contrary, for pure $e/p$ content,
the energetics relevant to thermal electrons 
and NT and thermal protons is all quite uncertain.
Based on Eq. (\ref{eq:p-rho}) and Table 1,
we can derive
\begin{eqnarray}
1<e
/e^{\rm NT}_{e}< 1\times10^{4} && ({\rm equal} ~ \rho) \nonumber \\ 
1<e
/e^{\rm NT}_{e}< 1\times10^{4} && ({\rm equal} ~ m)\nonumber \\ 
1<e
/e^{\rm NT}_{e}< 1\times10^{4} && ({\rm equal} ~ E)   .
\end{eqnarray}
For the case of maximum values of 
$\rho/\rho^{\rm NT}_{e}$ in Eq. (\ref{eq:p-rho}),
the total kinetic power for pure $e/p$ content reaches 
$L_{\rm kin}\sim 10^{4}L_{\rm kin,e}^{\rm NT}$
which is extremely large and unlikely.

\section{On the shock dissipation rate}\label{sec:dissipation}

In order to examine the allocation of 
the bulk kinetic energy of cold shells 
$E_{\rm blk}$ into the thermal energy,
we estimate the shock dissipation rate 
of bulk kinetic energy of colliding cold shells.
Here we denote the thermal energy of shocked shells as 
$E-E_{0}$ where $E_{0}$ is the rest mass energy
and $E$ is the total kinetic energy which satisfies 
$E\propto \Gamma_{2}e_{2}+\Gamma_{3}e_{3}$.
Then the shock dissipation rate 
$\epsilon_{\rm diss}$ defined as the ratio of 
the thermal energy of mass elements after the collision
to that of bulk kinetic energy
of mass elements before the collision 
is given by
\begin{eqnarray}\label{eq:diss} 
\epsilon_{\rm diss}
&\equiv& 
\frac{\Gamma_{2}(E-E_{\rm 0})}
{E_{\rm blk}-E_{\rm 0}} \nonumber \\
&=&
\frac{
 \Gamma_{2}
[(\Gamma_{12}-1)\delta m_{2}
+(\Gamma_{34}-1)\delta m_{3}]}
{(\Gamma_{1}-1)\delta m_{2}
+(\Gamma_{4}-1)\delta m_{3}} 
\end{eqnarray}
where
$\delta m_{2}$ and $\delta m_{3}$ are
the surface mass of the shocked regions 2 and 3, 
respectively.
Here, 
$\delta m_{2}$ and 
$\delta m_{3}$ are expressed as
$\delta m_{2}=\Gamma_{2}\rho_{2}(v_{\rm FS}-v_{\rm CD})\delta t$
and 
$\delta m_{3}=\Gamma_{3}\rho_{3}(v_{\rm CD}-v_{\rm RS})\delta t$
where $\delta t$ is the corresponding duration time 
in the ISM frame.
%
Two differences between the present work and 
the two-point-mass collision model (e. g., Piran 1999)
are that 
(i) we estimate $\epsilon_{\rm diss}$ 
with the shock junction conditions, and 
(ii) we subtract the irreducible rest mass term from 
the  denominator.
Our definition is superior to the previous 
one when the value of relative
Lorentz factor (i.e., $\Gamma_{34}$ and/or $\Gamma_{12}$)
are close to order unity and/or a small Lorentz factors
for cold shells.
From Eq. (\ref{eq:diss}) we obtain
\begin{eqnarray}
0.07 <
\epsilon_{\rm diss}< 0.44  && ({\rm equal} ~ \rho) \nonumber \\ 
0.09 <
\epsilon_{\rm diss}< 0.63   && ({\rm equal} ~ m) \nonumber \\ 
0.06 <
\epsilon_{\rm diss}<0.35  
&& ({\rm equal} ~ E)  .
\end{eqnarray}
The larger (smaller) $\Gamma_{\rm r}/\Gamma_{\rm s}$ becomes, 
the larger (smaller) $\epsilon_{\rm diss}$ realizes.
As previously mentioned (Kobayashi and Sari 2001;  KMY),
the case for equal mass of colliding shells 
realizes  largest value of $\epsilon_{\rm diss}$ and
an asymmetry of each mass reduces the 
value of $\epsilon_{\rm diss}$.

The fraction of
$1-\epsilon_{\rm diss}$ of the bulk kinetic energy
of the cold shells
survives and transferred to a larger scale. 
This is responsible for  the 
large scale  structure such as radio lobes and cocoons.
Therefore
the comparison with the large scale kinetic power
such as an extended radio emissions of blazars (e.g., 
Antonucci \& Ulvestad 1985)
will be an important future work although this is beyond the 
scope of this work.

\section{Summary and discussion}\label{sec:summary}

Invisible plasma content in blazar jets
such as protons and/or thermal $e^{\pm}$ pairs is investigated.
In this work,
we divide total mass and energy densities into 
two components, i. e., those of non-thermal electrons and those
of the other invisible components. 
It enable us to constrain on the amount of 
invisible plasma in the jet.
This is a significant forward step compared with previous studies.

The methodology of constraining 
the invisible plasma content in the emission region is as follows.
The lower limit to the mass and energy densities of total plasma
is limited by the definition that the mass and energy densities
of total plasma should be larger than those of 
the non-thermal electrons. 
The total mass and energy densities are
constrained by the internal shock dynamics. 
On the other hand, the upper limit of 
mass and energy densities for non-thermal electrons are 
constrained by the condition that bremsstrahlung emission from 
thermal electron (and positron) component should 
not exceed the observed SSC $\gamma$-ray emission. 
We can thus bracket the amount of total mass and energy densities
in the emission region from below and above.


We apply this method to the archetypal TeV blazar Mrk 421 
and obtain the following results.

\bigskip

(1) {\it Mildly relativistic shock is realized.}

By imposing the condition of the 
bulk Lorentz factor of the emission region as $\Gamma_{3}=12$
estimated by the multi-frequency spectrum of Mrk 421 (KTK),
we explore the allowed range of $\Gamma_{\rm r}/\Gamma_{\rm s}$
within the framework of the standard internal shock model.
Adopting the conditions of  
$\Gamma_{\rm s,min}>3$,
$\Gamma_{\rm r,max}<100$, and
$\Gamma_{\rm r}/\Gamma_{\rm s}>2$
based on the literatures 
(Wardle and Aarons 1997;
Begelman, Rees and Sikora 1994;
NP02),
we find that the values of $\Gamma_{\rm r}/\Gamma_{\rm s}$
for Mrk 421 are limited in the ranges of 
$2<\Gamma_{\rm r}/\Gamma_{\rm s}<16$ (equal $\rho$),
$2<\Gamma_{\rm r}/\Gamma_{\rm s}<19.5$ (equal $m$), and 
$2<\Gamma_{\rm r}/\Gamma_{\rm s}<11.7$ (equal $E$),
respectively.
As mentioned in Kirk and Duffy (1999), a very hard
injection index of $s\sim 1.6$ observed in Mrk 421 well agrees 
with this mildly relativistic 
shock regime (See Fig. 3 in their paper).  
Hence we conclude that mildly relativistic shocks
take place in Mrk 421 
from the analysis of 
the observed spectrum and the internal shock dynamics.

(2) 
{\it The mass density of invisible
plasma is much heavier than that of non-thermal electrons.}

Using the condition that the mass and energy densities
of non-thermal electrons should be lower than those
of the total ones, we derive the 
lower limit of total mass density at the shocked region.
Since the relative Lorentz factor between the shocked and unshocked regions
is expected to be a few (in Table 1), 
copious amount of mass density 
of invisible plasma is inevitably required.
The upper limit of  $n^{\rm T}_{e}$ is constrained by
the condition that the luminosity of bremsstrahlung emission 
should be smaller than the observed $\gamma$-ray luminosity
which is well explained by
the synchrotron-self-Compton emission.
Combining them, the allowed ranges of
$\rho/\rho_{e}^{\rm NT}$ 
for pure $e^{\pm}$ pair content are found as
$2\times 10^{2}<\rho/\rho^{\rm NT}_{e}<2\times 10^{3}({\rm equal} ~ \rho)$,
$7\times 10^{1}<\rho/\rho^{\rm NT}_{e}<2\times 10^{3}({\rm equal} ~ m)$, and  
$6\times 10^{1}<\rho/\rho^{\rm NT}_{e}<2\times 10^{3}({\rm equal} ~ E)$,
respectively.
For pure $e/p$ plasma content, 
the upper limit of $\rho/\rho_{e}^{\rm NT}$ turns out to be
$3\times 10^{6}$. 

Although the specific index $s=1.6$ for Mrk 421
is discussed here, we emphasize that the value $s<2$ is 
common character for TeV blazars as
they indeed display the smaller $s$ 
than 2 (Kirk and Duffy 1999 for review).
For instance, the choice of $s=2.2$ leads to the
synchrotron emission with $\nu F_{\nu}\propto \nu^{0.4}$.
Such a soft spectrum significantly conflicts 
with the observed synchrotron emission in blazars 
(e. g., Fossati et al. 1998).

(3) 
{\it  Electron acceleration efficiency in the shocked
region is evaluated.}

Once $\rho/\rho^{\rm NT}_{e}$ is bounded as shown 
in Figs. 1, 2, and 3, 
we can obtain the electron acceleration efficiency 
as
$e/e^{\rm NT}_{e}=
\Gamma_{34}\rho/\langle\gamma_{e}\rangle\rho^{\rm NT}_{e}$
for given $\Gamma_{34}$.
Since the allowed value of $\rho/\rho_{\rm NT}$ 
is in the narrow range for the case of pure $e^{\pm}$ content, 
we obtain the electron acceleration efficiency as
$1 < e/e^{\rm NT}_{e} < 7$.
Correspondingly 
the total kinetic power of the emission region $L_{\rm kin}$ 
resides in the range $1<L_{\rm kin}/L_{\rm kin,e}^{\rm NT}<7$.
It is clear that
a loading of proton component significantly
enlarges $L_{\rm kin}$. 
For $e/p$ plasma content,
the value  of $L_{\rm kin}$ is evaluated as 
$1<L_{\rm kin}/L_{\rm kin,e}^{\rm NT}<10^{4}$.
In the case of this maximal $L_{\rm kin}$, too large
$L_{\rm kin}$ could lead to a problem for the energy,
although we do not have a tight constraint 
on the amount of proton loading.

(4) 
{\it  The shock dissipation rate 
of bulk kinetic energy of colliding cold shells 
is examined.}

In \S 5, we further estimate the dissipation rate 
of bulk kinetic energy of colliding cold shells 
into the internal one via the shocks.
It is qualitatively clear that
the larger (smaller) $\Gamma_{\rm r}/\Gamma_{\rm s}$ becomes, 
the larger (smaller) $\epsilon_{\rm diss}$ realizes.
The resultant
shock dissipation rate for the colliding cold shells resides in
the range $\epsilon_{\rm diss}\sim 6 \textrm{--}60 \%$.

\bigskip

Lastly, let us  discuss 
the origin of blazar sequence (Fossati et al. 1998)
which is tightly connected to the nature of the central engine.
In Fossati et al. (1998)
they computed average spectral energy distributions from radio 
to gamma-rays for complete sample of blazars.
The resultant spectra show a continuity in that 
(i) the synchrotron peak occurs in different frequency for different 
samples/luminosity classes, with most luminous blazars peaking at lower 
frequencies; 
(ii) the peak frequency of the gamma-ray component 
correlates with the peak frequency of the lower energy one; 
(iii) the luminosity ratio between the high and low energy
components increases with bolometric luminosity. 
They claimed that the continuous sequence of properties  
may be controlled by a single parameter, related to 
the bolometric luminosity.
Below we enlighten another new ingredient for the origin of sequence.
Flat spectrum radio quasars
(FSRQs) have the order of magnitude larger kinetic power and 
energy densitity of the external radiation field 
than BL Lacs (e.g., Sikora et al. 1997). 
Hence, the leptonic components in FSRQs ejecta 
undergo stronger radiation drag effect 
in larger external radiation fields 
(Sikora and Wilson 1981; Phinney 1982;
see also Iwamoto and Takahara 2002).
However, the bulk Lorentz factors in FSRQs are 
comparable to or even slightly larger than the ones in TeV blazars
in spite of being subject to much stronger radiation drag
(e.g., Kubo et al. 1998; Spada et al. 2001; Kusunose et al. 2003).
In order to realize larger kinetic powers 
and larger bulk Lorentz factors against the strong radiation drag, 
we may take a new conjecture that 
a larger baryon loading may occur for FSRQs.
Summing up,  not only the strength of the external radiation field
but also the total amount and/or blending ratio 
of $e^{\pm}$ pair and $e/p$ could be new key 
quantities to explore the origin of the continuous blazar sequence.

\section*{Acknowledgments}
We thank the anonymous referees and H. Ito for invaluable comments.
We acknowledge the
Grant-in-Aid for Scientific Research 
of the Japanese Ministry of Education, Culture, Sports, Science
and Technology, No. 14079025, 14340066, and 16540215.






\end{document}